\documentclass[amsmath,amssymb,floats,preprint,preprintnumbers]{revtex4}
\usepackage{graphicx}% Include figure files
\usepackage{dcolumn}% Align table columns on decimal point
\usepackage{bm}% bold math

\def\be{\begin{equation}}
\def\ee{\end{equation}}
\def\bea{\begin{eqnarray}}
\def\eea{\end{eqnarray}}

\begin{document}
\title{Detecting Dichroism in Angle Resolved Photoemission}
\author{C. M. Varma}
\affiliation{Department of Physics, University of California,
  Riverside, CA 92521}

\begin{abstract}
Recently, the time-reversal violation predicted for the pseudogap phase of the cuprates, which was observed by dichroism experiments using Angle-Resolved Photoemission has also been observed by polarized neutron diffraction. Earlier derivation of dichroism in angle resolved photoemission due to time-reversal violation relied on existence of mirror planes in the crystal. Here the theory of the effect is generalized to the case that mirror plane symmetry is weakly violated due to perturbing potentials such as a superstructure.
\end{abstract}

\maketitle

\section{Introduction}

A theory for the pseudogap phenomena in the Cuprates suggests that it is a phase with broken time-reversal symmetry due to an ordered array of current loops, without change in translational symmetry \cite{cmv1,cmv2}. Such an elusive phase can be detected by dichroism in circularly polarized angle resolved photoemission \cite{cmv3, simon-varma}. Such experiments were done and found the effect \cite{kaminski}. Recently the broken symmetry deduced in such experiments has been directly observed by polarized neutron diffraction experiments \cite{bourges, mook}.

I first summarize the proposal for the dichroic experiment \cite{simon-varma}:  First, one must take note of the {\it geometric} dichroic effect which exists with and without any time-reversal breaking. This is the difference in intensity of outgoing photo-current for incident right circular polarized photons and left circular polarized photons if the momentum of the initial electronic state for photoemission is not in the mirror plane of the crystal while the  photon is incident in the mirror plane. This effect is null when the momentum of the initial state lies in the mirror plane and it is odd about variations of momentum perpendicular to it. In addition to this, if the crystal is in a phase which breaks time-reversal symmetry accompanied by reflection symmetry breaking about this mirror plane, the difference is non-zero at this mirror plane and is even for variations perpendicular to it. So on cooling from high temperatures where time-reversal is preserved to below a temperature at which time-reversal symmetry is broken, the momentum for which the dichroic effect is zero moves away from the mirror plane. This is the dichroic effect to be looked for to show time-reversal violation as well as to provide the detailed specification of the state through discovery of  the {\it crystalline} mirror planes about  which the reflection symmetry is lost.

It is important to realize (i) that both time-reversal and (some) mirrror plane symmetry must be lost to see the effect and (ii) the dichroic effect is observed because it is linear in the time-reversal violating wavefunction;  experiments sensitive only to
the $|wavefunction|^2$ (such as ordinary charge diffraction experiments or ARPES with linear polarization) are insensitive to the time-reversal violation as well as the accompanying  loss of reflection symmetry of the wave-function. The latter kind of experiment is of-course sensitive to loss of reflection symmetry arising from perturbations in atomic positions or potentials which lower the symmetry of the crystal. Proving time-reversal violation below a certain temperature therefore requires the occurrence of the dichroic effect below that temperature unaccompanied by a loss of symmetry in the latter type of experiment.

A legitimate question can be raised about the dichroism experiments performed: the theory of the effect, as summarized above relies on the existence of mirror planes in the crystal. But due to the superstructure modulation, a buckling of the $BiO$ layer, no mirror plane strictly exists on the surface of the BISCCO crystals on which the ARPES experiments are performed. How then to interpret the observed results? It ought to be mentioned that since the surface reconstruction is observed not to be temperature dependent and the dichroic effect is zero above a certain temperature and changes its magnitude below it, prima facie, the dichroism cannot be due to this loss of symmetry. However, it is worth investigating how the dichroism effect may be modified due to loss of mirror symmetry.
This is the aim of this addenda.

The conclusion is that provided the loss of mirror plane symmetry is due to a small perturbation $\epsilon \ll 1$ on the electronic states studied in ARPES, the zero of the dichroic effect shifts linearly in $\epsilon$, while a finite effect due to time-reversal violation occurs at the position where the geometric effect is zero and is even about it. So, the point of zero of the dichroic effect moves on breaking time-reversal symmetry just as it does when there is a {\it crystalline} mirror plane. This conclusion is derived below.

\section{Geometric and time-reversal induced dichroism in the absence of a mirror plane} 

 Suppose a beam of circularly polarized photons of energy $\omega$ shone on a crystal in the
  direction $\hat{n}$ produces  free-electons with momentum $\bf {p}$
  and energy $E_ {\bf {p}}$ at the detector. Let $|\bf{k}\rangle$  denote the
  states of the crystal. Here $\bf{k}$ is the wave-vector in the first
  Brillouin zone. Also $\bf{k}=\bf{p}$, modulo the reciprocal vectors, because the photon is assumed to have negligible momenta compared to the electrons. Quite generally, 
\bea
 |\bf{p}\rangle = \alpha_m |{\bf{p}},e\rangle + \beta_m|{\bf{p}},o\rangle; \\
 |{\bf{k}} \rangle = \mu_m |{\bf{k}},e \rangle + \nu_m |{\bf{k}},o \rangle .
 \eea
 The division of the wavefunctions into two parts is made such that in reflection on {\it any} plane the part labeled $\bf{e}$ does not change sign and the part labeled $\bf{o}$ does:
 \bea
  \Re_m |{\bf{p}}\rangle = \alpha_m |{\bf{p}},e\rangle - \beta_m |{\bf{p}},o\rangle; \\
   \Re_m|{\bf{k}} \rangle = \mu_m |{\bf{k}},e \rangle -  \nu_m|{\bf{k}},o \rangle.
   \eea
   
   If the reflection were on a mirror plane and $\bf{p},\bf{k}$ were in the mirror plane, $\beta_m=\nu_m=0$. This was the case treated in (\onlinecite{simon-varma}). Suppose we consider a plane $m$ which looses the  symmetry of a mirror plane weakly due to a perturbing potential. Then for $\bf{p},\bf{k}$ in $m$, $\beta_m$ and $\nu_m$ are of $O(\epsilon) \ll 1$. We will also consider further small deviations, also of $O(\epsilon)$ from $\bf{p},\bf{k}$ in $m$.
   
   The matrix element for photemission is
   \be
    \langle {\bf p |M_i| k}\rangle = \frac{-ie}{2mc} \langle{\bf{p}}|
     {\bf{A_i.\nabla}} |{\bf{k}}\rangle,
     \ee
where the subscript $i$ refers to circular polarization, either right ($r$) or left ($l$).
 For simplicity I will consider the situation when the direction $\hat{n}$ of the incident photon is normal to the crystal surface. Then
   \be
    \Re^{-1}_m(\bf{A_{\ell}.\nabla})\Re_m = (\bf{A_r.\nabla})
    \ee
    continues to hold.  
    
    As in Ref.(\cite{simon-varma}), the difference in the outgoing current $J_{\bf p}$ for the two polarizations is calculated from the difference in the (squared) matrix element
    ${\cal D}_m$. For $\epsilon,\epsilon'\ll1$, so that $\beta_m, \nu_m \ll 1$, we need only keep the first and the third terms of Eq.(8) of Ref.(\onlinecite{simon-varma}), so that
    \be
    {\cal{D}}_m \approx  
        4 {\cal{R}} \left( \alpha_m^*\beta_m |\mu_m|^2  \langle {\bf{p}},e|{\bf M}_r^*|{\bf{k}},e \rangle
\langle {\bf{k}},e|{\bf M}_r |{\bf{p}},o \rangle +  
     \mu_m\nu_m^*|\alpha_m|^2\langle {\bf{p}},e|{\bf M}_r^*|{\bf{k}},e\rangle 
          \langle {\bf{k}},o|{\bf M}_r|{\bf{p}},e\rangle \right) .
 \ee
  Now, unlike in Ref.(\cite{simon-varma}), ${\cal{D}}_m \ne 0$, but of $O(\epsilon)$, when
 $\bf{p},\bf{k}$ are in $m$. However, suppose $\bf{p},\bf{k}$ are moved by $\delta\bf{p}_{\perp}, \delta\bf{k}_{\perp}$ orthogonal to $m$ so that they lie in a plane $m'$. Then there is an additional contribution to 
 $ {\cal{D}}_m$, which is proportional to $\delta\bf{p}_{\perp}$ and $\delta\bf{k}_{\perp}$, because this part of the wavefunction must change sign for
 $(\delta\bf{p}_{\perp}, \delta\bf{k}_{\perp}) \to (-\delta\bf{p}_{\perp}, -\delta\bf{k}_{\perp})$. (Due to loss of mirror plane, this statement is true only to $O(\epsilon)$; this makes the conclusions arrived at here violated to $O(\epsilon^2)$). It therefore follows
that by adjusting the magnitude of $\delta\bf{p}_{\perp}/\bf{p}$ and $\delta\bf{k}_{\perp}/\bf{k}$ to $O(\epsilon)$ of one or the other sign, the geometric effect can be brought to $0$.

Now, let us consider the effect of time-reversal violation on ${\cal{D}}_m$. The experiment to look for the dichroism effect due to time-reversal violation is done after locating $m'$ as described above. Due to time-reversal violation and attendant {\it extra} loss of mirror plane symmetry, the coefficients $\beta_m, \nu_m$ acquire extra amplitudes proportional to the time-reversal breaking order parameter for ${\bf p,k}$  in $m'$. Further, again to $O(\epsilon)$, this effect is even for further small further variations of $\delta{\bf p'}_{\perp},\delta{\bf k'}_{\perp}$ about either side of $m'$. 

To summarize: The conditions for the experiment if there are  perturbations leading to a weak loss of mirror symmetry are  to find incident ${\bf k}$ in $m'$ at which the geometric effect is zero and such that for small perpendicular variations about about such a ${\bf k}$, the effect is odd. This effect should be temperature independent and always present. Then cool below the symmetry breaking temperature and observe a temperature dependent effect which is even for small perpendicular variations of ${\bf k}$ about $m'$.

\section{Experiments}

Actually,  the experiment which saw the effect \cite{kaminski} was done precisely in the manner specified above. A  geometric effect was found above a certain temperature $T_g$ in each sample (within the uncertainties of the experiments, it coincided with the temperature at which the pseudogap begins to be observed in resistivity and thermodynamic experiments). The zero of the geometric effect, which is independent of T for $T\gtrsim T_g$ was located. Below $T_g$, the dichroic effect was observed in that the zero of dichroism moved as a function of temperature.

Thus in the experiments, the considerations given here to account for any small loss of {\it crystalline} mirror plane symmetry were (unwittingly) taken care of. In reality the effect of the super-structure perturbation is so weak that that their effect is invisible in any loss of symmetry in linearly polarized ARPES from states from which the dichroic effect is observed. It was estimated \cite{kaminski2} that the effect must be less than $0.1\%$, while the dichroic effect observed is 1 to 2 $\%$. Besides, even this small superstructure is not known to have sensitivity to temperature in crystallographic measurements.

Thus there are two different ways that the experiment done is free of the criticizms made on the basis of the superstructure \cite{armitage, borisenko, golden}. First as already noted \cite{kaminski2}, the effect of superstructure is too small and is temperature independent. Second, even if it were significant (but still small), the slightly generalized theory of the effect presented above shows that the procedure followed in the experiment allows deduction of dichroism. Recent direct observation \cite{bourges, mook}
through polarized neutron diffraction further settle the issue of the broken symmetry in underdoped cuprates.

The calculation presented here may be useful in situations in which crystalline mirror plane symmetry is broken on the same scale as the expected dichroic effect.

 \end{document}